\documentclass[letterpaper,english,reprint,nofootinbib,aps,superscriptaddress,apl]{revtex4-1}

\usepackage{babel,calc,amsmath,amsthm,amssymb,graphicx,,subfigure,xcolor,comment}

\usepackage{array}
\usepackage{txfonts}
\usepackage{amsmath}
\usepackage{graphicx}
\usepackage{dcolumn}
\usepackage{bm}
\usepackage{ulem}
\usepackage{mathdots}
\usepackage{todonotes}
\usepackage[unicode=true]{hyperref}
\usepackage{threeparttable}
\usepackage{times}
\usepackage{CJK}
\hypersetup{
	colorlinks=true,       		% false: boxed links; true: colored links
	linkcolor=blue,          	% color of internal links
	citecolor=blue,            % color of links to bibliography
	urlcolor=black,           	% color of external links
}

\begin{document}

\begin{CJK*}{GB}{gbsn}

\title{Third-harmonic generation of spatially structured light in a quasi-periodically poled crystal}

%%%%%%%%%%%%%%%%%%%%%%%%%%%%%
\author{Yan-Chao Lou}
\author{Zi-Mo Cheng}
\author{Zhi-Hong Liu}
\author{Yu-Xiang Yang}
\author{Zhi-Cheng Ren}
\author{Jianping Ding}
\author{Xi-Lin Wang}
\email[]{xilinwang@nju.edu.cn}	
\affiliation{National Laboratory of Solid State Microstructures and School of Physics, Nanjing University, Nanjing 210093, China}
\affiliation{Collaborative Innovation Center of Advanced Microstructures, Nanjing University, Nanjing 210093, China}

\author{Hui-Tian Wang}
\email[]{htwang@nju.edu.cn}
\affiliation{National Laboratory of Solid State Microstructures and School of Physics, Nanjing University, Nanjing 210093, China}
\affiliation{Collaborative Innovation Center of Advanced Microstructures, Nanjing University, Nanjing 210093, China}

\date{\today}

%%%%%%%%%%%%%%%%%%%%%%%%%%%%%%%%%%%

\begin{abstract}

Nonlinear optical processes of spatially structured light, including optical vortex and vector optical fields, have stimulated a lot of interesting physical effects and found a variety of important applications ranging from optical imaging to quantum information processing. However, high harmonic generation of vector optical fields with space-varying polarization states is still a challenge. Here we demonstrate third harmonic generation of spatially structured light including vector optical fields, in a nonlinear Sagnac interferometer containing a carefully designed quasi-periodically poled potassium titanyl phosphate for the first time. The experimental results are in good agreement with the theoretical predictions. Our results will enable to manipulate spatially structured light or photons at new wavelengths and carrying higher orbital angular momentum. Our approach has the potential applications for the research of optical skyrmions and may open up new opportunities to produce spatially structured entangled photons for quantum communication and computation. 
\end{abstract}
	
\maketitle

\end{CJK*}

Spatially structured light~\cite{Rubinsztein-Dunlop2017, Wang2020, Forbes2021} has attracted broad interest because of its novel effects and important applications. For example, the subluminal effect has been observed~\cite{Giovannini2015}. Various novel optical fields could be created, such as optical needle~\cite{Wang2008}, optical polarization M\"{o}bius strips~\cite{Bauer2015}, and spatiotemporal optical vortices~\cite{Chong2020}. Spatially structured light has been found useful in many important applications from classical optics~\cite{Andrews2008} to quantum optics~\cite{Fobres2019}.   

Optical nonlinearity of spatially structured light has become an important and hot topic in recent years. As a typical spatially structured light, an optical vortex with helical phase front of $\exp (j m \varphi)$ (where $\varphi$ is the azimuthal angle in polar coordinate system and $m$ is the topological charge) could carry an orbital angular momentum (OAM) of $m\hbar$ per photon~\cite{Allen1992, Padgett2017}. A variety of nonlinear interactions involving OAM have been demonstrated, including second harmonic generation (SHG)~\cite{Dholakia1996, Bloch2012, Shemer2013, Li2013, Zhou2014, Chaitanya2015, Ni2016, Chen2020, Tang2020, Gui2021, Hancock2021}, third harmonic generation (THG)~\cite{Fang2016, Xu2018}, high-order harmonic generation~\cite{Gauthier2017}, and spontaneous parametric down-conversion (SPDC)~\cite{Mair2001, Leach2010, Dada2011}. The introduction of spatial phase structure in nonlinear interaction will result in novel effects and promote new applications. For instance, besides the energy and linear momentum conservations in nonlinear process, the OAM conservation~\cite{Dholakia1996, Mair2001, Leach2010, Dada2011, Li2013, Fang2016, Xu2018, Gui2021, Hancock2021} has been explored in many nonlinear interactions involving optical vortices. The applications of nonlinear interaction with OAMs are prolific in many realms such as optical imaging~\cite{Qiu2018}, holography~\cite{Fang2021}, and high-dimensional quantum information processing~\cite{Erhard2020}. 

Vector optical fields, as another typical spatially structured light, have space-varying polarization structures. Such kind of spatially structured light can manipulate more photonic degrees of freedom simultaneously and then lead to a variety of novel functionalities, such as spin-orbit photonics~\cite{Cardano2015}, alignment-free communication~\cite{Aolita2007, DAmbrosio2012}, and enhancing channel capacity~\cite{Barreiro2008, Zhu2021}. It is more challenging to explore the nonlinear process of vector optical fields, which involves various polarization components, while the phase-matching condition generally restricts nonlinear interaction only for a fixed combination of polarization states. The SHG of vector optical fields has been reported by a nonlinear interferometer~\cite{YangOL2019} or two orthogonally configured nonlinear crystals~\cite{LiuOL2018} to achieve the nonlinear interaction between two components of vector optical field. However, the experimental realization of THG of vector optical field is still a challenge.

In this letter, we demonstrate the first THG of vector optical field in a nonlinear Sagnac interferometer with a quasi-periodically poled potassium titanyl phosphate (QPPKTP). The quasi-periodic structure of second-order nonlinear coefficient can provide more abundant reciprocal vectors, which is a flexible approach for achieving multiple second-order nonlinear interactions in a single crystal. For example, the THG was demonstrated experimentally in one-dimensional nonlinear quasicrystals~\cite{Zhu1997, Zhang2001, Vernay2020} and a theoretical design of two-dimensional photonic quasicrystals for multiple nonlinear frequency conversion was proposed~\cite{Lifshitz2005}. Here, the single QPPKTP produces third harmonic (TH) wave by implementing two second-order nonlinear optical processes simultaneously: the SHG of a fundamental wave (FW), and the sum-frequency generation (SFG) of the FW and the second harmonic (SH) wave.

Here we devote to the THG of vector optical field. To achieve our goal, a good choice is to use a QPPKTP, which is designed by following the projection method~\cite{Zhang2001, Zia1985}. To use the largest second-order nonlinear coefficient, the QPPKTP is designed as a type-0 phase-matching scheme, i.e., all the interacting waves are vertically polarized (denoted by $| V \rangle$). As shown in Fig.~1(a), our QPPKTP consists of two building blocks A and B. Each block (A or B) contains a pair of antiparallel domains. The width of A (B) is $l_A \! = \! l_{A^{+}} + l_{A^{-}}$ ($l_B \! = \! l_{B^{+}} + l_{B^{-}}$). The widths of domains, the projection angle $\theta$ (or $\gamma \! = \! \tan \theta$), and the arrangement of the QPPKTP (with a length of 5.532 mm) are listed in Fig.~1(a) 

\begin{figure}[!ht]
	\centering
	\includegraphics[width=0.85\linewidth]{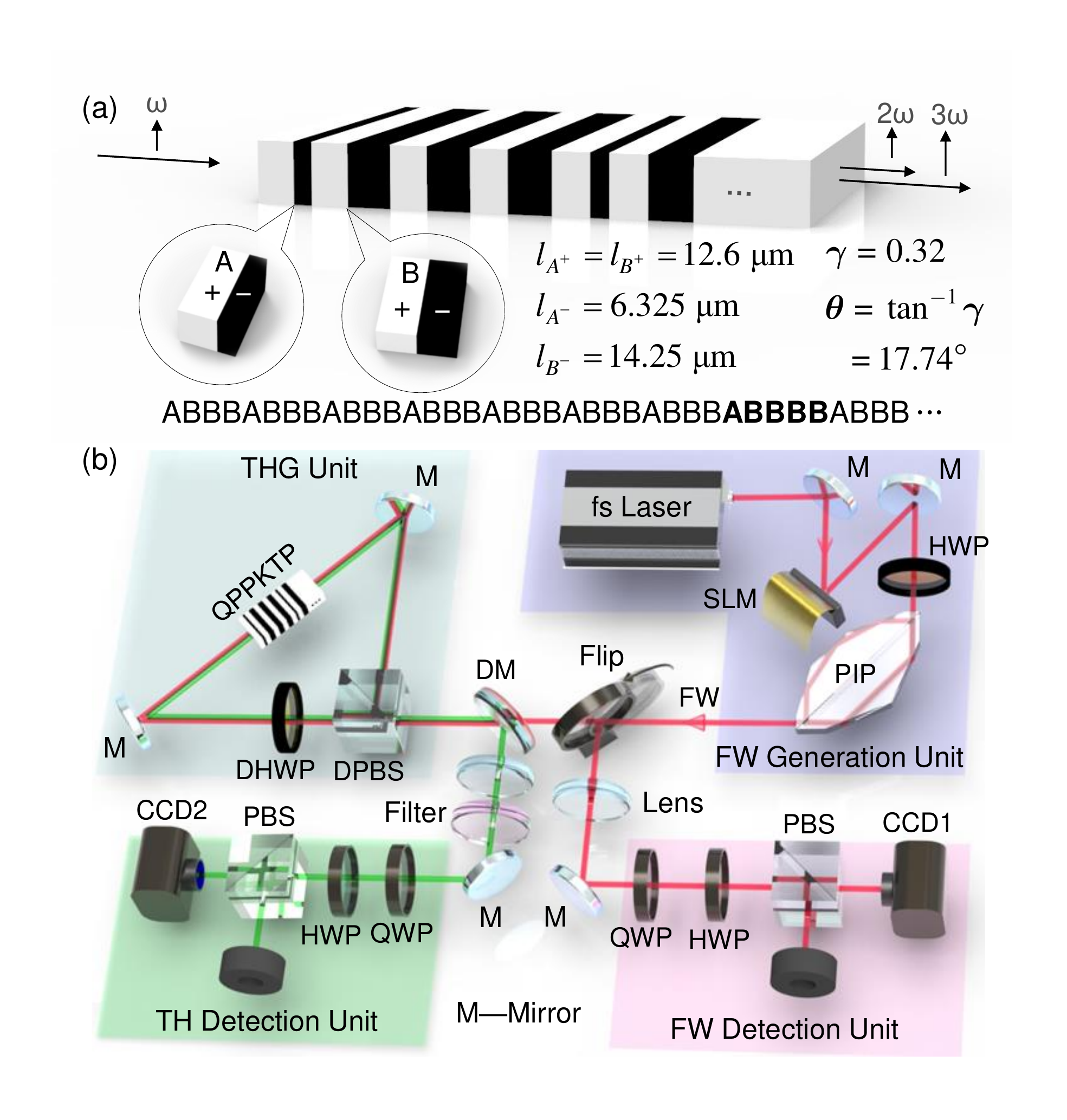}
	\caption{{\small THG of vector optical field. (a) QPPKTP consists of two building blocks A and B, and each block contains a pair of antiparallel domains. (b) Configuration for THG of vector field in a double-pass nonlinear Sagnac interferometer with a QPPKTP. There are four units: the generation and detection units of the FW vector field, the THG unit, the TH detection unit. Filter is used to block the SH at 780 nm.}}
\end{figure}

Under the non-depletion approximation and the phase-matching conditions, when the input FW field is a $| V \rangle$-polarized vortex with a topological charge of $m$ (its state is written as $| V \rangle | m \rangle$), the generated $| V \rangle$-polarized TH field in the type-0 QPPKTP should be written as $| V \rangle | 3m \rangle$~\cite{Zhu1997}. Clearly, the topological charge of the TH vortex field is triple that of the FW field because three FW photons are converted into one TH photon and the OAM is conserved in this THG process~\cite{Fang2016, Xu2018}. Our QPPKTP is only valid for the $| V \rangle$-polarized case. For instance, for a FW vector field containing both $| V \rangle$- and $| H \rangle$-polarized (horizontally polarized) vortices, its state can be written as
\begin{equation}
| \phi \rangle =\alpha| H \rangle | \! + \! m \rangle + \beta| V \rangle | \! - \! m \rangle,
\end{equation}
where $\alpha$ and $\beta$ are the normalized factors with $\alpha^2 + \beta^2 = 1$. For convenience, we choose the two vector field expression bases as $| H \rangle$ and $| V \rangle$, which are the two bases for the vector field generation element utilized in our experiment: polarization interferometric prism (PIP)~\cite{Ren2021}. A single-pass configuration with only one QPPKTP cannot realize the THG of the two components for a vector field, hence a double-pass configuration should be a good solution. Here we utilize a Sagnac interferometer with the QPPKTP shown in Fig.~1(b) to implement a double-pass THG. For the FW vector field, its $| H \rangle$-polarized ($| V \rangle$-polarized) component is transmitted (reflected) by a polarizing beam splitter (PBS), and then propagates along the clockwise (counterclockwise) loop. It is of great importance to insert a half-wave plate (HWP) at 45$^\circ$ into the Sagnac interferometer, which converts the $| H \rangle$-polarized component into the $| V \rangle$-polarized one to generate the $| V \rangle$-polarized TH field. As PBS and HWP in the Sagnac interferometer should also work for the TH field, both are designed to be dual-wavelength PBS (DPBS) and dual-wavelength HWP (DHWP). And reflection will invert the sign of OAM or topological charge of the optical vortex~\cite{Ren2021}, suggesting that the odd (even) reflections will invert (keep) the sign of topological charge of the optical vortex.

In the counterclockwise loop, the $| V \rangle$-polarized FW vortex undergoes only two reflections (by the DPBS and a mirror) and then  generates the $| V \rangle$-polarized TH vortex in the QPPKTP; after one reflection (by a mirror) and passing through the DHWP at $45^\circ$, the TH vortex becomes $| H \rangle$-polarized and its sign of OAM is also reversed; finally it outputs from the Sagnac interferometer through the DPBS. In the clockwise loop, the $| H \rangle$-polarized FW vortex is converted to be $| V \rangle$-polarized by the DHWP at $45^\circ$; after one reflection by a mirror (the sign of OAM is reversed), it is incident into the QPPKTP; the generated $| V \rangle$-polarized TH vortex experiences two reflections (by a mirror and the DPBS) and then outputs from the Sagnac interferometer. As a result, the two orthogonally polarized components of the TH fields generated in the two loops are combined at the DPBS to generate the TH vector field propagating along the direction opposite to the FW one. As stated above, therefore, when the FW vector field shown in Eq. (1) is incident into the Sagnac interferometer, correspondingly, the output TH vector field from the THG unit should have the following form of state
\begin{align}
| \phi' \rangle = (\alpha^6 + \beta^6)^{-1/2} (\alpha^3 | V \rangle | \! - \! 3m \rangle + \beta^3 | H \rangle | \! + \! 3m \rangle).
\end{align} 
Clearly, the $| H \rangle$-polarized ($| V \rangle$-polarized) FW vortex at the state $ | H \rangle | \! + \! m \rangle$ ($ | V \rangle | \! - \! m \rangle$) generates the $| V \rangle$-polarized ($| H \rangle$-polarized) TH vortex at the state $ | V \rangle | \! - \! 3m \rangle$ ($ | H \rangle | \! + \! 3m \rangle$) from the Sagnac interferometer. The generated TH vector field is reflected by a dichroic mirror (DM) to be separated from the FW one and then enters the TH detection unit in Fig.~1(b).

In experiment, we can generate the FW vector fields by the FW generation unit shown in Fig.~1(b). A femtosecond (fs) laser (FF ULTRA 1560, TOPTICA Photonics Inc.) with a central wavelength of 1560 nm, a repetition frequency of 80 MHz, and a pulse duration of ~200 fs is incident on a spatial light modulator (SLM, model P1920, Meadowlark Optics Inc.). The SLM is loaded the programmable holographic grating to produce the FW vortex field with variable topological charge, which is incident into PIP after passing through a HWP, finally the FW vector fields can be generated (the details can be found in Ref.~\cite{Ren2021}). Of course, our FW generation unit generates easily the scalar optical vortices with homogeneous polarization. It should be pointed out that the profiles of the FW fields we generated are very similar to the zero-radial-index Laguerre-Gaussian (LG) beams.

Firstly, we would like to demonstrate the THG of scalar FW optical vortex in the state of $| P \rangle | m \rangle$, where $| P \rangle$ denotes the polarization state, such as $| H \rangle$-polarized ($| V \rangle$-polarized) state and right-handed (left-handed) circularly polarized state $| R \rangle \! = \! (| H \rangle - j | V \rangle) / \sqrt{2} $ ($ | L \rangle \! = \! (| H \rangle + j | V \rangle) / \sqrt{2}$). Here we prepare three FW optical vortices in the states of $| V \rangle | \! - \! 1 \rangle$, $| L \rangle | \! - \! 2 \rangle$ and $| H \rangle | \! - \! 3 \rangle$. The $| H \rangle$- or $| V \rangle$-polarized FW optical vortex will propagates along the clockwise or counterclockwise loop of the nonlinear Sagnac interferometer independently, while the circularly polarized FW optical vortex will pass through both the clockwise and counterclockwise loops simultaneously. 

To detect the FW optical vortices, a flip mirror is utilized to reflect it from the main optical path and another mirror is utilized to preserve its topological charge before entering the FW detection unit shown in Fig.~1(b). The intensity patterns of the FW vortices at 1560 nm are measured by CCD1 (SP907-1550, Ophir-Spiricon Inc.). The experimentally measured patterns of the FW vortex fields with the states of $| V \rangle | \! - \! 1 \rangle$, $| L \rangle | \! - \! 2 \rangle$ and $| H \rangle | \! - \! 3 \rangle$ are shown in the first, third and fifth columns of Fig.~2. CCD2 (LBP2-HR-VIS2, Newport Corporation) is used to record the intensity patterns of the TH fields at 520 nm by using the TH detection unit shown in Fig.~1(b). The experimentally measured patterns of the TH vortices with theoretically predicted states of $| H \rangle | \! + \! 3 \rangle$, $| R \rangle | \! + \! 6 \rangle$ and $| V \rangle | \! + \! 9 \rangle$ are shown in the second, fourth and sixth columns of Fig.~2, respectively. The topological charges are measured by a tilt lens (the last row of Fig.~2), in which the stripe patterns along the diagonal or anti-diagonal direction indicate the positive or negative topological charge, and the number of dark stripes give the absolute value of the topological charge~\cite{VaityPLA2013}. Clearly, the polarization states and topological charges of the TH vortex fields are in good agreement as predicted by theory as shown in Eq.~(2). 

\begin{figure}[!ht]
	\centering
	\includegraphics[width=0.85\linewidth]{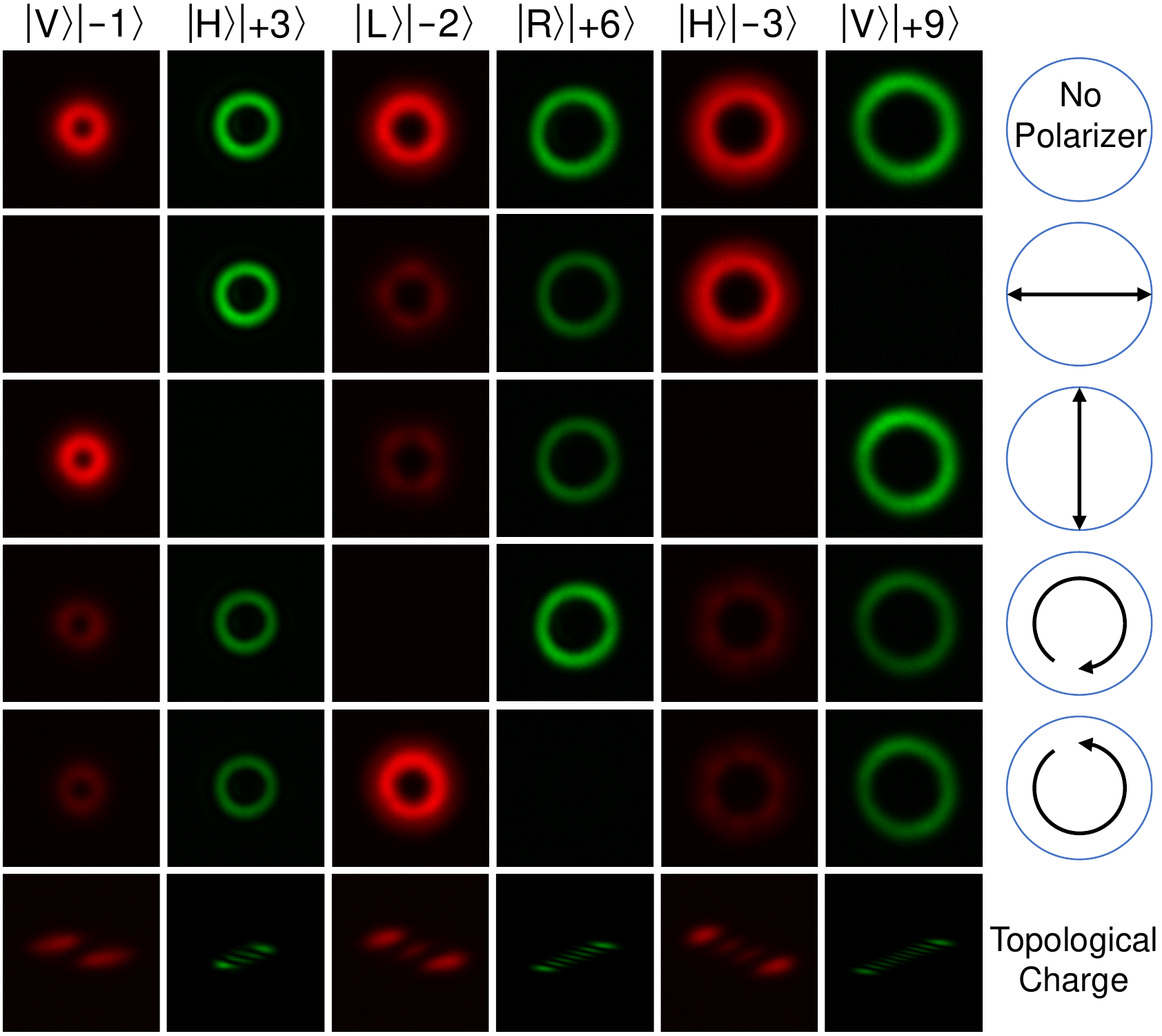}
	\caption{{\small Experimental results for THG of scalar vortices. All the red patterns show the experimental results for the scalar FW vortices, while all the green patterns show the corresponding TH results. First row shows the total intensity patterns for the FW and TH scalar vortices. Second-fifth rows are the measured intensity patterns of the components under the bases of $| H \rangle$, $| V \rangle$, $| R \rangle$ and $| L \rangle$. The last row shows the patterns of the measured vortices after passing through a tilt lens to reveal topological charges.}}
\end{figure} 

Next, we would like to investigate the THG of vector fields. Using the FW generation unit in Fig.~1(b), we prepare the FW vector fields in the three states 
\begin{align}
| \phi_1 \rangle & = 2^{- 1/2} \left(| H \rangle | \! - \! 1 \rangle + | V \rangle | \! + \! 1 \rangle \right), \\
| \phi_2 \rangle & = 2^{- 1/2} \left(| H \rangle | \! - \! 4 \rangle - | V \rangle | \! + \! 4 \rangle \right), \\
| \phi_3 \rangle & = \sin 35^\circ | H \rangle | \! - \! 1 \rangle + \cos 35^\circ | V \rangle | \! + \! 1 \rangle.
\end{align}
With Eq.~(2), the three FW vector fields in Eqs.~(3)-(5) will generate the corresponding TH fields with the following states 
\begin{align}
| \phi'_1 \rangle & = 2^{- 1/2} \left(| V \rangle | \! + \! 3 \rangle + | H \rangle | \! - \! 3 \rangle \right), \\
| \phi'_2 \rangle & = 2^{- 1/2} \left(| V \rangle | \! + \! 12 \rangle - | H \rangle | \! - \! 12\rangle \right), \\
| \phi'_3 \rangle & = 1.721 (\sin^3 35^\circ | V \rangle | \! + \! 3 \rangle + \cos^3 35^\circ | H \rangle | \! - \! 3 \rangle).
\end{align} 

To evaluate the qualities of the prepared FW vector fields and the generated TH vector fields, we will use the Stokes parameters normalized by the total intensity, as $S_1 \! = \! E_x E^\ast_x \! - \! E_y E^\ast_y$, $S_2 \! = \! E_x E^\ast_y \! + \! E^\ast_x E_y$ and $S_3 \! = \! j (E_x E^\ast_y \! - \! E^\ast_x E_y)$~\cite{Born1999,Wang2010}, where $E_x$ and $E_y$ indicate the normalized complex amplitudes of the $| H \rangle$- and $| V \rangle$-polarized components, respectively. We measure  the total intensity patterns and the intensity patterns under the polarization bases of $| H \rangle$, $| V \rangle$, $| D \rangle \! = \! (| H \rangle - | V \rangle) / \sqrt{2}$, $| A \rangle \! = \! (| H \rangle + | V \rangle) / \sqrt{2}$, $| R \rangle$ and $| L \rangle$. The second-fourth rows of Fig.~3 show the normalized $S_1$, $S_2$ and $S_3$ calculated from the corresponding patterns measured above. Clearly, all the experimental results of three prepared FW vector fields and their TH vector fields are in good agreement with our expectations in Eqs.~(3)-(8). 

\begin{figure}[!ht]
	\centering
	\includegraphics[width=0.95\linewidth]{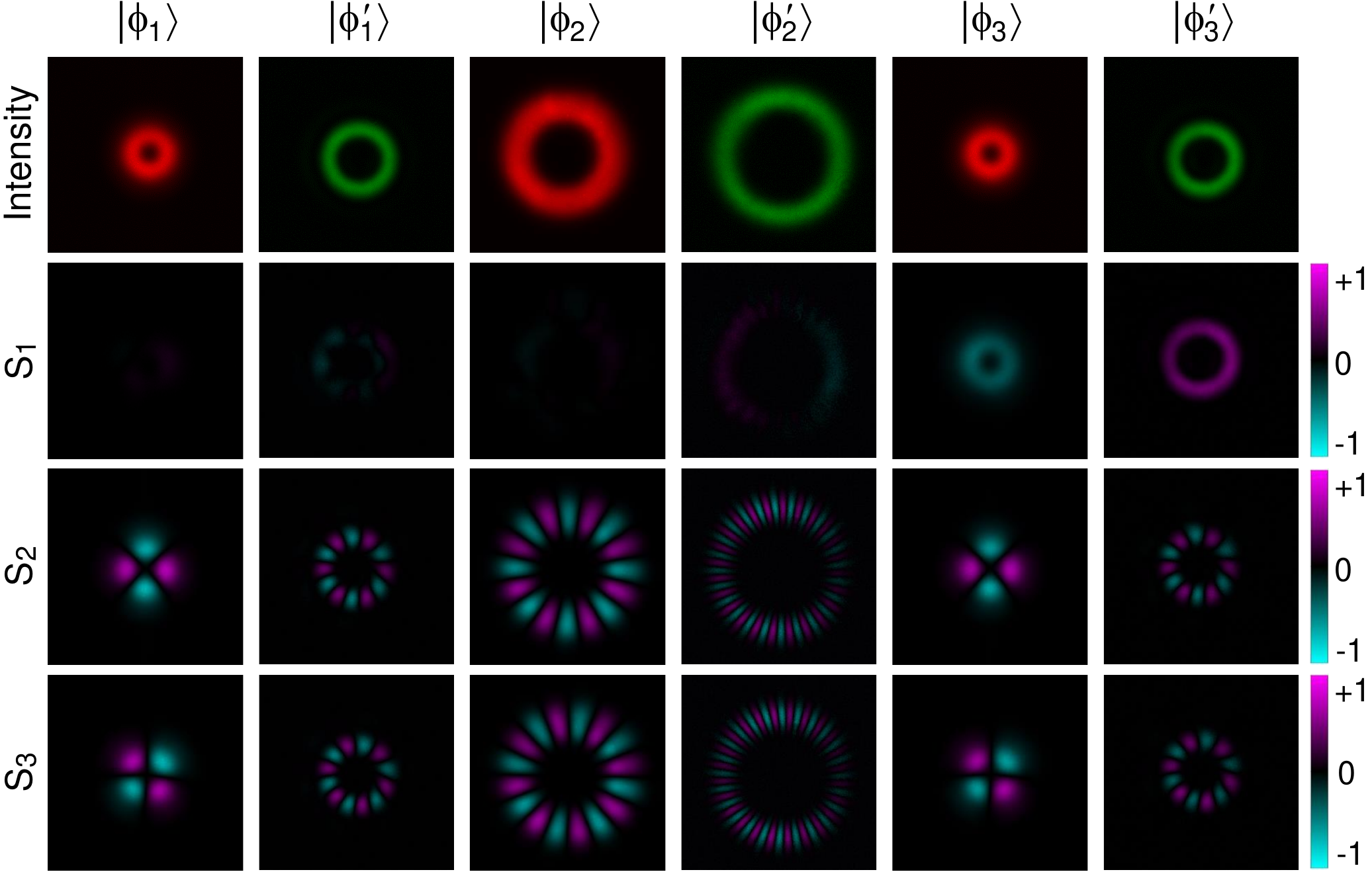}
	\caption{{\small Experimentally measured total intensity patterns and the calculated normalized Stokes parameters with values ranging from $-1$ to $+1$ by the difference of the measured intensity patterns for the prepared three FW vector fields (first, third and fifth columns) and their TH vector fields (second, fourth and sixth columns).}}
\end{figure}

As for the conversion efficiency of our QPPKTP, when the FW field is a fundamental Gaussian mode with an average power of 2 W and a radius of 100 $\mu$m, the SHG efficiency is 17.6$\%$ and the THG efficiency is 8.0$\%$, suggesting that the QPPKTP we designed and prepared should be efficient. However, the power of the generated FW vector fields is limited by the damage threshold of the SLM in the near infrared band and the FW vector fields have an average pump power of 91.2 mW in our experiment. In addition, to achieve the high-quality TH vector fields, the FW vector fields are weakly focused into the QPPKTP, resulting in the lower power density, hence the THG efficiency are relatively low to be 1.2$\times$10$^{-4}$, 3.0$\times$10$^{-5}$, 1.0$\times$10$^{-4}$ and 3.1$\times$10$^{-5}$ for the FW fields of $| V \rangle | \! - \! 1 \rangle$, $| L \rangle | \! - \! 2 \rangle$, $| H \rangle | \! - \! 3 \rangle$ and $| \phi_1 \rangle $ in Eq.~(3), respectively. The THG efficiencies of the two FW fields of $| L \rangle | \! - \! 2 \rangle$ and $| \phi_1 \rangle $ are $\sim$1/4 that of the other two FW fields, as the reasons stated below. Differently from the THGs of $| V \rangle | \! - \! 1 \rangle$ and $| H \rangle | \! - \! 3 \rangle$ in only one loop, the two FW $| L \rangle | \! - \! 2 \rangle$ and $| \phi_1 \rangle $ have to be divided into the two loops with a half of power to realize the THGs, resulting in that the THG efficiency drops to $\sim$1/8 in one loop and the sum of THG efficiency in the two loops becomes $\sim$1/4. In future, the THG efficiency could be increased by increasing the pump power density through preparing high-power FW vector field and more tightly focusing. If the dual-wavelength elements are replaced by triple-wavelength elements, the experimental setup can output both the SH wave and TH wave of spatially structured light simultaneously.

In summary, we successfully demonstrated the THG of spatially structured light with only one QPPKTP, and in future, generalizing nonlinear crystal design to 3 dimensions may enable to generate SH and TH spatially structured light by using 3D nonlinear crystals~\cite{Zhang2021} instead of shaping the FW with an SLM. By using a nonlinear Sagnac interferometer, we realized the THG of vector fields with space-varying polarization states for the first time. Our approach may find potential applications to investigate optical skyrmions~\cite{Gao2020, Karnieli2021} by exploring pseudo-spin textures of vector fields in nonlinear optics. The inverse process, SPDC, is expected to produce spatially structured entangled photons and may open up new opportunity for high-dimension~\cite{Mair2001, Dada2011, Erhard2020} quantum information processing.

~\

\noindent \textbf{Funding.} {\small National Natural Science Foundation of China (No. 11922406 and No. 91750202); National Key R\&D Program of China (No. 2019YFA0308700, No. 2020YFA0309500, and No. 2018YFA0306200); Key R\&D Program of Guangdong Province (No. 2020B0303010001).}

\noindent \textbf{Acknowledgment.} {\small The authors would like to thank the support by Collaborative Innovation Center of Extreme Optics.}


\begin{thebibliography}{10}

	\bibitem{Rubinsztein-Dunlop2017} H. Rubinsztein-Dunlop, A. Forbes, M. V. Berry, M. R. Dennis, D. L. Andrews, M. Mansuripur, C. Denz, C. Alpmann, P. Banzer, T. Bauer, E. Karimi, L. Marrucci, M. J. Padgett, M. Ritsch-Marte, N. M. Litchinitser, N. P. Bigelow, C. Rosales-Guzm\'{a}n, A. Belmonte, J. P. Torres, T. W. Neely, M. Baker, R. Gordon, A. B. Stilgoe, J. Romero, A. G. White, R. Fickler, A. E. Willner, G. Xie, B. J. McMorran, and A. M. Weiner, J. Opt. \textbf{19}, 013001 (2017).
	
	\bibitem{Wang2020} J. Wang, F. Castellucci, and S. Franke-Arnold, AVS Quantum Sci. \textbf{2}, 031702 (2020).
	
	\bibitem{Forbes2021} A. Forbes, M. Oliveira, and M. R. Dennis, Nature Photon. \textbf{15}, 253 (2021). 
	
	\bibitem{Giovannini2015} D. Giovannini, J. Romero, V. Potocek, G. Ferenczi, F. Speirits, S. M. Barnett, D. Faccio, and M. J. Padgett, Science \textbf{347}, 857 (2015).
	
	\bibitem{Wang2008} H. Wang, L. Shi, B. Lukyanchuk, C. Sheppard, and C. T. Chong, Nature Photon. \textbf{2}, 501 (2008).    
	
	\bibitem{Bauer2015} T. Bauer, P. Banzer, E. Karimi, S. Orlov, A. Rubano, L. Marrucci, E. Santamato, R. W. Boyd, and G. Leuchs, Science \textbf{347}, 964 (2015).
	
	\bibitem{Chong2020} A. Chong, C. Wan, J. Chen, and Q. Zhan, Nature Photon. \textbf{14}, 350 (2020). 
	
	\bibitem{Andrews2008} D. L. Andrews, \textit{Structured light and its applications: an introduction to phase-structured beams and nanoscale optical forces} (Academic Press-Elsevier, Burlington, 2008).
	
	\bibitem{Fobres2019} A. Forbes and I. Nape, AVS Quantum Sci. \textbf{1}, 011701 (2019).
	
	\bibitem{Allen1992} L. Allen, M. W. Beijersbergen, R. J. C. Spreeuw, and J. P. Woerdman, Phys. Rev. A. \textbf{45}, 8185 (1992).
	
	\bibitem{Padgett2017} M. J. Padgett, Opt. Express \textbf{25}, 11265 (2017).
	
	\bibitem{Dholakia1996} K. Dholakia, N. B. Simpson, M. J. Padgett, and L. Allen, Phys. Rev. A \textbf{54}, R3742 (1996).
	
	\bibitem{Bloch2012} N. Voloch-Bloch, K. Shemer, A. Shapira, R. Shiloh, I. Juwiler, and A. Arie, Phys. Rev. Lett. \textbf{108}, 233902 (2012).
	
	\bibitem{Shemer2013} K. Shemer, N. Voloch-Bloch, A. Shapira, A. Libster, I. Juwiler, and A. Arie, Opt. Lett. \textbf{38}, 5470 (2013).
	
	\bibitem{Li2013} S. M Li, L. J Kong, Z. C. Ren, Y. N. Li, C. H. Tu, and H. T. Wang, Phys. Rev. A. \textbf{88}, 035801 (2013).
	
	\bibitem{Zhou2014} Z. Y. Zhou, Y. Li, D.-S. Ding, W. Zhang, S. Shi,  B. S. Shi, and G. C. Guo, Opt. Express \textbf{22}, 23673 (2014).
	
	\bibitem{Chaitanya2015} N. A. Chaitanya, A. Aadhi, M. V. Jabir, and G. K. Samanta, Opt. Lett. \textbf{40}, 2614 (2015).
	
	\bibitem{Ni2016} R. Ni, Y. F. Niu, L. Du, X. P. Hu, Y. Zhang, and S. N. Zhu, Appl. Phys. Lett. \textbf{109}, 151103 (2016).
	
	\bibitem{Chen2020} Y. Chen, R. Ni, Y. Wu, L. Du, X. Hu, D. Wei, Y. Zhang, and S. N. Zhu, Phys. Rev. Lett. \textbf{125}, 143901 (2020).
	
	\bibitem{Tang2020} Y. Tang, K. Li, X. Zhang, J. Deng, G. Li, and E. Brasselet, Nature Photon. \textbf{14} 658 (2020).
	
	\bibitem{Gui2021} G. Gui, N. J. Brooks, H. C. Kapteyn, M. M. Murnane, and C.-T. Liao, Nature Photon. \textbf{14}, 658 (2021).
	
	\bibitem{Hancock2021} S. W. Hancock, S. Zahedpour, and H. M. Milchberg, Optica \textbf{8}, 594 (2021). 
	
	\bibitem{Fang2016} X. Y. Fang, G. Yang, D. Z. Wei, D. Wei, R. Ni, W. Ji, Y. Zhang, X. P. Hu, W. Hu, Y. Q. Lu, S. N. Zhu, and M. Xiao, Opt. Lett. \textbf{41}, 1169 (2016).
	
	\bibitem{Xu2018} Z. Xu, Z. Y. Lin, Z. L. Ye, Y. Chen, X. P. Hu, Y. D. Wu, Y. Zhang, P. Chen, W. Hu, Y. Q. Lu, M. Xiao, S. N. Zhu, Opt. Express \textbf{26}, 17563 (2018).
	
	\bibitem{Gauthier2017} D. Gauthier, P. R. Ribic, G. Adhikary,A. Camper,C. Chappuis, R. Cucini, L. F. DiMauro,G. Dovillaire,F. Frassetto, R. Geneaux, P. Miotti ,L. Poletto, B. Ressel, C. Spezzani, M. Stupar, T. Ruchon, and G. De Ninno, Nature Commun. \textbf{8}, 14971 (2017).
	
	\bibitem{Mair2001} A. Mair, A. Vaziri, G. Weihs, and A. Zeilinger, Nature \textbf{412}, 313 (2001).
	
	\bibitem{Leach2010} J. Leach, B. Jack, J. Romero, A. M. Jha, A. M. Yao, S. Franke-Arnold, D. G. Ireland, R. W. Boyd, S. M. Barnett, and M. J. Padgett, Science \textbf{329}, 662 (2010).
	
	\bibitem{Dada2011} A. C. Dada, J. Leach,G. S. Buller, M. J. Padgett, and E. Andersson, Nat. Phys. \textbf{7}, 677 (2011).
		
	\bibitem{Qiu2018}  X. Qiu, F. Li, W. Zhang, Z. Zhu, and L. Chen, Optica \textbf{5}, 208 (2018).
	
	\bibitem{Fang2021} X. Y. Fang, H. C. Yang, W. Z. Yao, T. X. Wang, Y. Zhang, M. Gu, and M. Xiao, Adv. Photon. \textbf{3}, 015001 (2021).
	
	\bibitem{Erhard2020} M. Erhard, M. Krenn, and A. Zeilinger, Nature Rev. Phys. \textbf{2}, 365 (2020).
	
	\bibitem{Cardano2015} F. Cardano and L. Marrucci, Nature Photon. \textbf{9}, 776 (2015).
	
	\bibitem{Aolita2007} L. Aolita and S. P. Walborn, Phys. Rev. Lett. \textbf{98}, 100501 (2007).
	
	\bibitem{DAmbrosio2012} V. D'Ambrosio, E. Nagali, S. P. Walborn, L. Aolita, S. Slussarenko, L. Marrucci, and F. Sciarrino, Nature Commun. \textbf{3}, 961 (2012).
	
	\bibitem{Barreiro2008} J. T. Barreiro, T.-C Wei, and P. G. Kwiat, Nature Phys. \textbf{4}, 282 (2008). 
	
	\bibitem{Zhu2021} Z. Y. Zhu, M. Janasik, A. Fyffe, D. Hay, Y. Y. Zhou, B. Kantor, T. Winder, R. W. Boyd, G. Leuchs, and Z. M. Shi, Nature Commun. \textbf{12}, 1666 (2021).
	
	\bibitem{YangOL2019} C. Yang, Z. Y. Zhou, Y. Li, Y. H. Li, S. L. Liu, S. K. Liu, Z. H. Xu, G. C. Guo, and B. S. Shi, Opt. Lett. \textbf{44}, 219 (2019).
	
	\bibitem{LiuOL2018} H. G. Liu, H. Li, Y. L. Zheng, and X. F. Chen, Opt. Lett. \textbf{43}, 5981 (2018).
	
	\bibitem{Zhu1997} S. N. Zhu, Y. Y. Zhu, and N. B. Ming, Science \textbf{278}, 843 (1997).
		
	\bibitem{Zhang2001} C. Zhang, H. Wei, Y. Y. Zhu, H. T. Wang, S. N. Zhu and N. B. Ming, Opt. Lett. \textbf{26}, 899 (2001).
	
	\bibitem{Vernay2020} A. Vernay, L. Bonnet-Gamard, V. Boutou, S. Trajtenberg-Mills, A. Arie, and B. Boulanger, OSA Continuum. \textbf{3}, 1536 (2020).
	
	\bibitem{Lifshitz2005} R. Lifshitz, A. Arie, and A. Bahabad, Phys. Rev. Lett. \textbf{95}, 133901 (2005).
	
	\bibitem{Zia1985} R. K. P. Zia and W. J. Dallas, J. Phys. A \textbf{18}, L341 (1985).
	
	\bibitem{Ren2021} Z. C. Ren, Z. M. Cheng, X. L. Wang, J. P. Ding and H. T. Wang, Appl. Phys. Lett. \textbf{118}, 011105 (2021).
	
	\bibitem{VaityPLA2013} P. Vaity, J. Banerji, and R. P. Singh, Phys. Lett. A \textbf{377}, 1154 (2013).
	
	\bibitem{Born1999} M. Born and E. Wolf, Principles of Optics, 7th ed. (Cambridge U. Press, 1999).
	
	\bibitem{Wang2010} X. L. Wang, Y. Li, J. Chen, C. S. Guo, J. Ding, and H. T. Wang, Opt. Express \textbf{18}, 10786 (2010).
	
	\bibitem{Zhang2021} Y. Zhang, Y. Sheng, S. N. Zhu, M. Xiao, and W. Krolikowski, Optica \textbf{8}, 372 (2021).
	
	\bibitem{Gao2020} S. J Gao, F. C. Speirits, F. Castellucci, S. Franke-Arnold, S. M. Barnett, and J. B. G\"{o}tte, Phys. Rev. A \textbf{102}, 053513(2020).
	
	\bibitem{Karnieli2021} A. Karnieli, S. Tsesses, G. Bartal, and A. Arie, Nature Commun. \textbf{12}, 1092 (2021).
	
\end{thebibliography}
\end{document}